%% file: prcCosPA-okuda.tex
\documentclass[12pt]{article}
\usepackage{graphicx}

\newcommand\pubnumber{}
\newcommand\pubdate{}

\def\email{\footnote{okudat@fc.ritsumei.ac.jp, okuda@icrr.u-tokyo.ac.jp}}
					
\def\Title#1{\begin{center} {\Large #1 } \end{center}}
\def\Author#1{\begin{center}{ \sc #1} \end{center}}
\def\Address#1{\begin{center}{ \it #1} \end{center}}

\newcommand\pubblock{\rightline{\begin{tabular}{l} \pubnumber\\
         \pubdate  \end{tabular}}}
\newenvironment{Abstract}{\begin{quotation}  }{\end{quotation}}
\newenvironment{Presented}{\begin{quotation} \begin{center} 
             PRESENTED AT\end{center}\bigskip 
      \begin{center}\begin{large}}{\end{large}\end{center} \end{quotation}}
\def\Acknowledgements{\bigskip  \bigskip \begin{center} \begin{large}
             \bf ACKNOWLEDGEMENTS \end{large}\end{center}}

\input econfmacros.tex

\begin{document}
\begin{titlepage}
\pubblock

\vfill
\Title{Recent results from Telescope Array}
\vfill
\Author{Takeshi Okuda\email \ for Telescope Array Collaboration}
\Address{Department of Physical Sciences, Ritsumeikan University\\
Kusatsu, Shiga, JAPAN}
\vfill
\begin{Abstract}
The Telescope Array (TA) observatory, located in midwest Utah, USA, is
designed to detect ultra high energy cosmic rays whose energy is greater than 
1 EeV. TA mainly consists of two types of detector. The first type is the
atmospheric Fluorescence Detector (FD). TA's three FDs  have been in
operation since Fall 2007. The other type of detector is a ground-covering
Surface Detector (SD), which has been operating at TA since Spring 2008. In
addition, the TA-Rader (TARA) and EUSO-TA associated experiments are
co-located with TA, and the TA Low Energy (TALE) extension recently started
partial operation. I report some recent general results from TA, and
describe our future plans.
\end{Abstract}
\vfill
\begin{Presented}
Symposium on Cosmology and Particle Astrophysics\\
(CosPA2013)
\end{Presented}
\vfill
\end{titlepage}
\def\thefootnote{\fnsymbol{footnote}}
\setcounter{footnote}{0}

\section{Introduction}

The flux of cosmic rays whose energies are over $10^{19}$eV, 
is about 1 particle/km$^2$/year.
Therefore, very large detector is 
required to detect these ultra high energy cosmic rays.

\begin{figure}[ht]
\centering
\includegraphics[width=10cm]{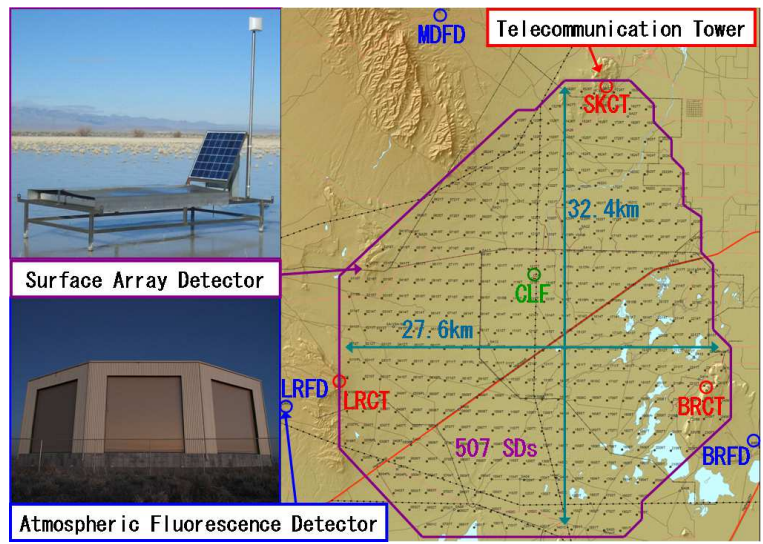}
\caption{TA site over view.}
\label{fig:TApic}
\end{figure}

The Telescope Array experiment (TA) is the observatory for air-shower caused 
by cosmic ray whose energy is greater than $10^{18.5}$eV.
This experiment takes in Millard County, Utah, United States 
(39.3$^\circ$N, 112.9$^\circ$W, altitude 1382 m).
Major construction of the TA started in 2005. 
The TA consists of 3 atmospheric fluorescence detector(FD) stations,
507 surface array particle detectors(SD), the central laser facility, 
the electron light source and 3 telecommunication towers, shown in  
Figure~\ref{fig:TApic}.

There are 12 FD telescopes in each station at Black Rock and Long Ridge 
(BRFD and LRFD). 
The FD telescope for BRFD and LRFD is 3.3 m diameter spherical mirror telescope 
which is divided by 18 segment mirrors.
The camera of telescope consists 256 PMTs whose field of view is 
$18^\circ$ in azimuth and $15^\circ$ in elevation.
One station covers $3^\circ\sim 33^\circ$ in elevation 
and $108^\circ$ in azimuth by 6$\times$2 telescopes \cite{NIMFD}.
The observation started in November 2007.

In addition, there are 14 telescopes at Middle Drum (MDFD), 
which come from HiRes-I \cite{HiRes}.
The FD telescope for MDFD is 2 m diameter mirror telescope 
which is divided by 4 segment mirrors. 
The camera of telescope consists of 256 PMTs whose field of view is 
$16^\circ$ in azimuth and $14^\circ$ in elevation. 
This station covers $3^\circ\sim 31^\circ$ in elevation 
and $114^\circ$ in azimuth by 7$\times$2 telescopes \cite{MDFD}.
The observation started in December 2007.

There is the central laser facility (CLF) at the center of the TA site, 
almost equidistant from 3 FD stations.
CLF YAG laser fires 100 shots twice an hour. 
This light is used as a standard light source for each FD stations and 
can be used to measure the atmospheric transparency 
which changes in a short time.
The observation started in December 2008.
In addition, there is a vertical LIDAR system by two PMTs for 
different height range \cite{CLF}.

There is the electron linear accelerator (ELS) at BRFD.
The accelerated electrons are bent to vertical in 100 m front of the telescope. 
This electron beam pulse is used to directly calibrate 
fluorescence telescopes and atmospheric fluorescence efficiency. 
The beam pulse is about 40 MeV $10^9$ electrons at 100 m from FD, 
which is scaled to the air-shower of 100 EeV at 10 km from FD \cite{ELS}.

There are 507 SDs, deployed in a square grid of 1.2 km spacing, 
covering 680 km$^2$.
An SD is 2 layers 3 m$^2$ plastic scintillation detector.
The thickness of one layer is 1.2 cm.
The scintillation light is collected to PMTs for each layer,
by wavelength shifting fibers.
The SD array is divided by 3 sub-arrays and the data collection 
takes place through 3 telecommunication towers.
The observation started in March 2008 and 
the cross boundary trigger for 3 sub-arrays 
was installed at the end of 2008 \cite{NIMSD}.

\section{Observational Results}
\subsection{Energy Spectrum}

\begin{figure}[ht]
\centering
\includegraphics[width=10cm]{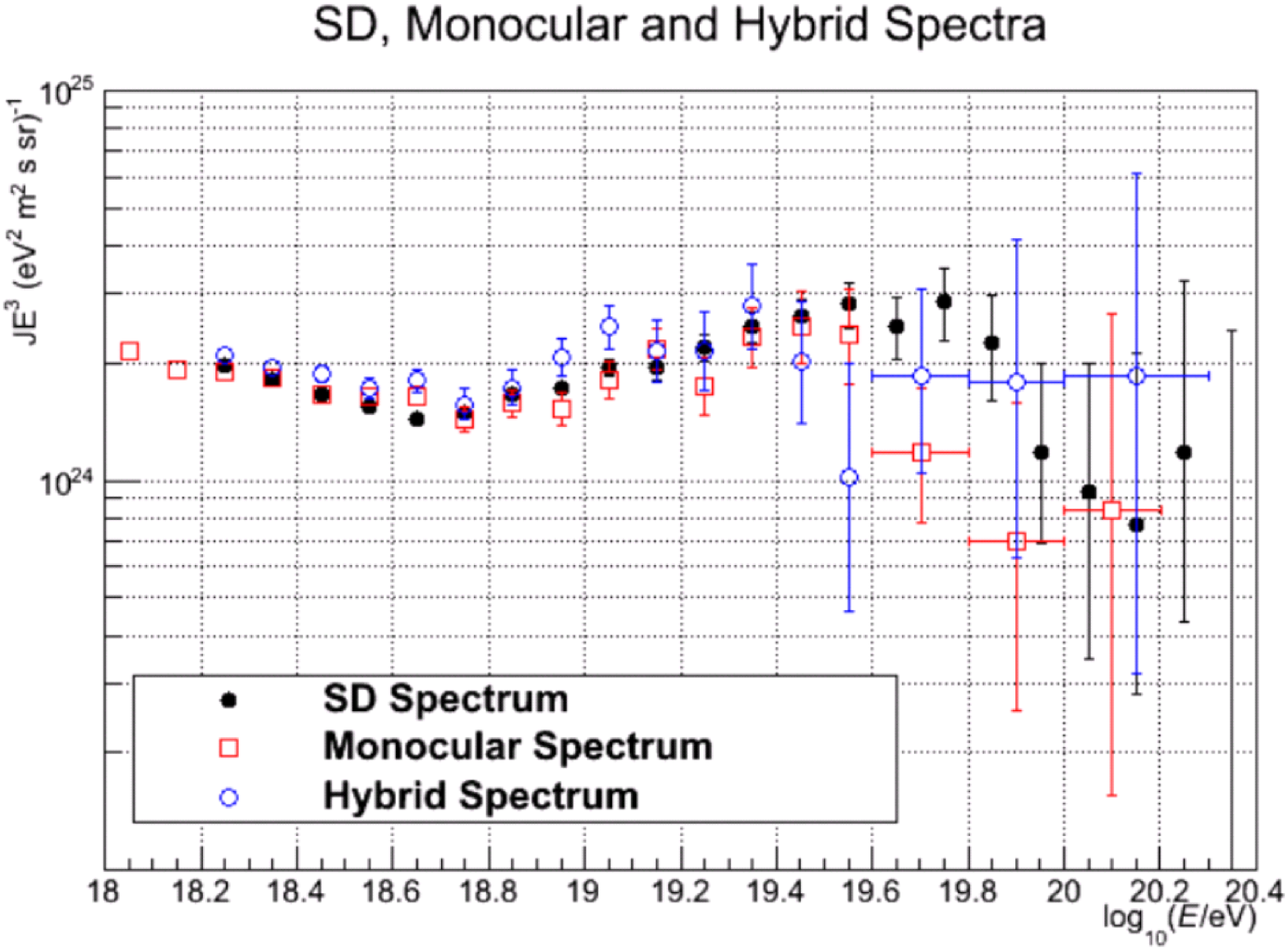}
\caption{Energy Spectrum from SD, monocular FD and hybrid data \cite{TAspec}.}
\label{fig:TAspec}
\end{figure}

There are three energy spectra by SD, monocular FD and hybrid observation 
\cite{TAspec}, shown in Figure~\ref{fig:TAspec}.

The SD energy spectrum depends on the Monte Carlo (MC) air-shower simulation.
We employ the value S(800) to estimate the primary cosmic ray energy.
S(800) is the fitted lateral particle density 
at 800 m away from the shower axis.
MC generates the look up table between energy and S(800) for each zenith angle.
This look up table energy is rescaled downward 27\% by FD energy measurement.
The SD effective aperture are also calculated by MC.
Five-years SD spectrum can be fitted 2-breaks broken power law \cite{TAspec}.
The two breaks of energy correspond to the features known as 
the "Ankle" and the "GZK Cutoff".
They are $E_{ankle}=5.05\pm0.27$ EeV and $E_{ankle}=56.8\pm10.5$ EeV, 
respectively.
The significance of the break at the GZK Cutoff is 5.7$\sigma$
with 68.1 events expected above the cutoff while only 26 events were observed.

On the other hand, FD energy estimation is calorimetric method
by fitting shower longitudinal development, 
employing certain static atmospheric transparency.
The monocular FD effective aperture is also calculated by MC.
The monocular FD spectrum is made by more statistics at lower energy,
and less statistics at higher energy than SD spectrum.
Therefore, it is difficult to compare SD and FD spectrum at GZK region,
but the ankle structure agrees well.

\subsection{Mass Composition}

There are two composition study by stereo FD and hybrid observation
\cite{STcomp,MDcomp}.

Since the air-shower developments have large fluctuation, it is
not easy to determine the species of the primary particle individually. 
Thus, the mass composition is determined
statistically, by comparing average, RMS or distribution of
$X_{max}$ of observed data with a MC simulation. 
$X_{max}$ is the slant column density of atmosphere 
where the number of secondary charged particles is maximum,
subtracting the slant column density where the shower starts. 
It should be noted that the uncertainty of the MC
depends strongly on hadronic interaction models that have
been extrapolated from measured cross sections at much lower energies.
As energy increases, the $X_{max}$ of air-showers increase.
And at a given energy, the $X_{max}$ of a light primary particles
will be deeper than that of a heavier primary particles. 

By stereo FD $X_{max}$ distribution of each energy bin, 
the hypothesis that the primary particles 
are 100\% iron is rejected for the energy than 10$^{19.4}$eV
by KS test in several hadronic interaction model.
Whereas, the hypothesis that the primary particle is 100\% proton
is consistent for the energy less than 10$^{19.4}$eV \cite{STcomp}.
For higher energy region, both 100\% proton and 100\% iron hypothesis
are consistent because of less statistics of observed data.

\begin{figure}[ht]
\centering
\includegraphics[width=10cm]{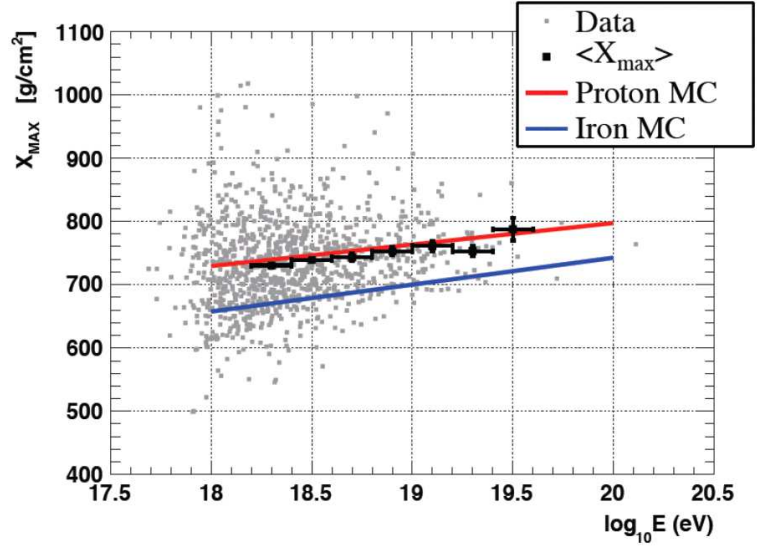}
\caption{$X_{max}$ from four-years MD hybrid data \cite{MDcomp}.
The average MC $X_{max}$ line is generated by QGSJET-II model 
for Proton and Iron as primary particles. The black data points with error 
bars represent the average data $X_{max}$, binned by energy.}
\label{fig:MDcomp}
\end{figure}

Figure~\ref{fig:MDcomp} shows the average $X_{max}$ by hybrid observation 
for each energy bin \cite{MDcomp}.
The observed average $X_{max}$ is consistent with 100\% proton case.

\subsection{Photon and Neutrino Search}

These study is done by inclined showers observed by SD.

The photon fraction study is done by using shower front curvature.
The air-shower initiated by photon starts development at lower sky
compared with hadron shower, statistically. 
Therefore, the shower front initiated by photon is curved more
than it by hadron, relatively.
We got the upper limit of photon flux using five-years data \cite{Photon}.

The Neutrino search is done by using very inclined shower waveforms.
The showers which started far from SD array give narrow one-peak waveform
for each detector. 
On the other hand, The showers which started near SD array give wide
indented multi-peak waveform for each detector. 
Therefore, the very inclined event which has wide
indented multi-peak waveform is candidate of the shower by neutrino.
We got the upper limit of neutrino flux using five-years data \cite{Photon}.

\subsection{Anisotropy}

There are several studies of the cosmic ray anisotropy by SD observation 
\cite{TAaniso}. 

\begin{figure}[ht]
\centering
\includegraphics[width=8cm]{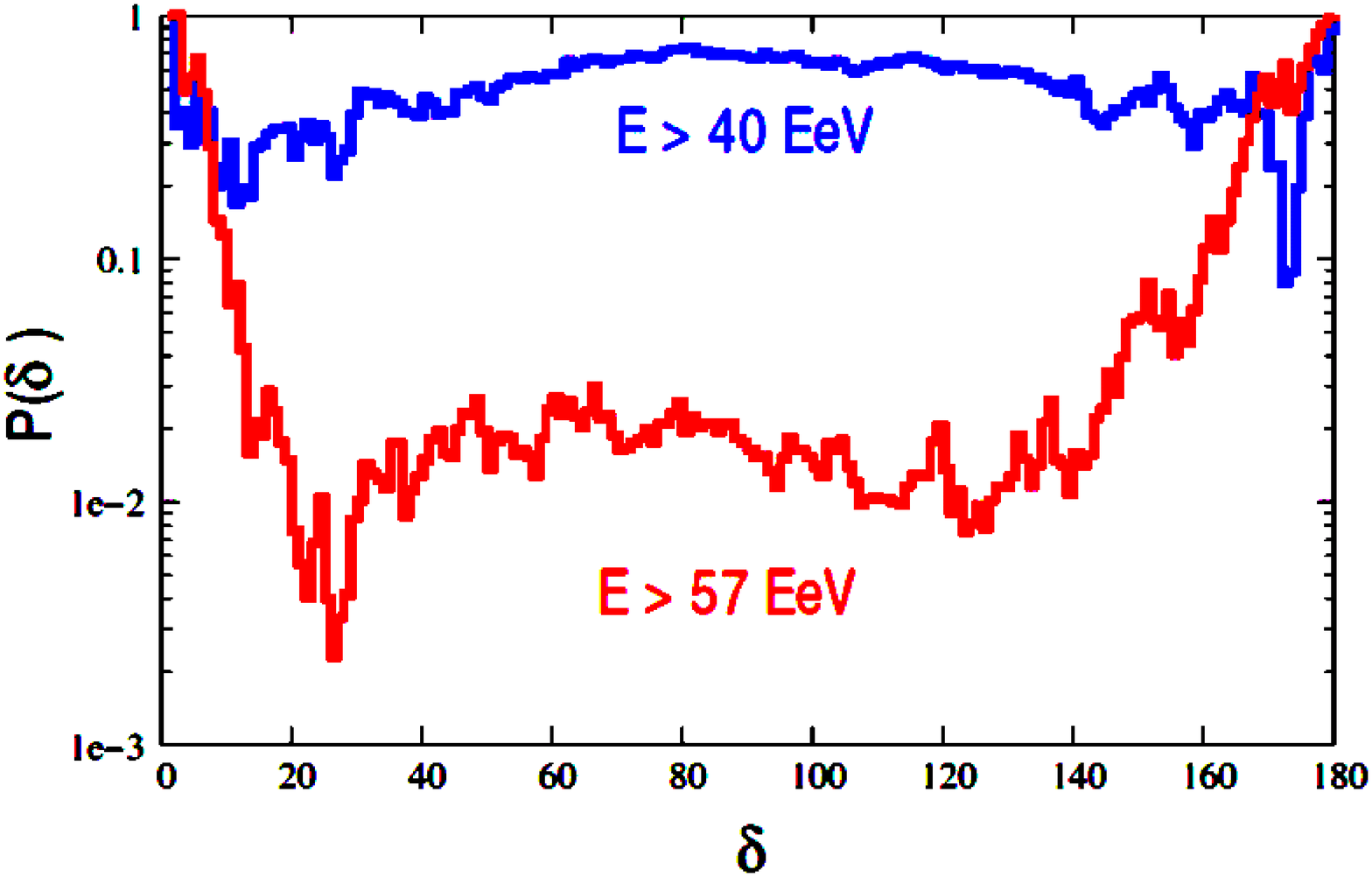}
\includegraphics[width=7cm]{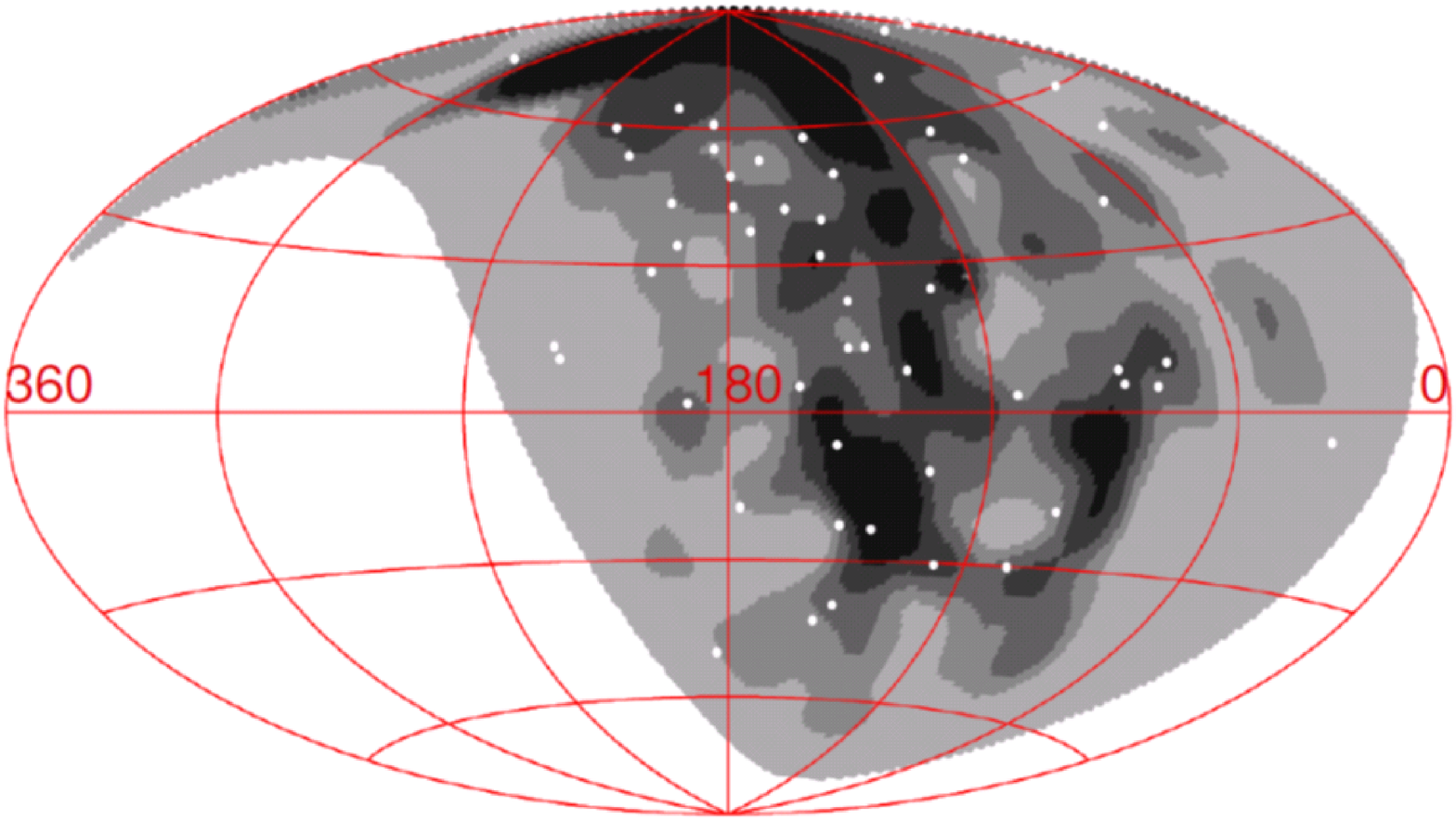}
\caption{(Left) The dependence of the p-value on the pair
separation angle for two energy thresholds. 
(Right) Sky map of the events with energy greater than 57 EeV 
in the Galactic coordinates. 
Bands of gray show the flux distribution expected in the 
LSS model of sources with the smearing angle of 6$^\circ$. \cite{TAaniso}.}
\label{fig:SDaniso}
\end{figure}

The result of cluster search by auto-correlation function is 
shown in the left of Figure~\ref{fig:SDaniso}.
The procedure is as follows: 
1st, setting energy threshold 
(i.e. setting the number of event for this evaluation).
2nd, calculating the opening angle for all pairs of observed event 
whose energy is over threshold.
3rd, counting the number of pairs whose opening angle is less than $\delta$.
4th, repeating 2nd and 3rd for a large number of simulated random event sets,
one set has same number of event as observed event set.
5th, calculating the fraction of simulated sets whose number of pairs 
(less than $\delta$) is larger than, or equal to, 
the number of pairs for observed event set.
This fraction is p-value, which describes how likely
the excess of pairs occurs by the fluctuation in random sets,
for certain $\delta$.
The left of Figure~\ref{fig:SDaniso} shows p-value for all $\delta$.
There is no excess of small-scale clusters in the TA data \cite{TAaniso}.

The result of AGN correlation was done by VCV catalog(2006) 
or other catalogs.
For the case using VCV catalog(2006),
we employed same parameter condition for precedent 
southern sky observation by Pierre Auger \cite{PAaniso}, to northern sky.
The p-value of correlation with AGN for this case is 0.013 \cite{TAaniso}.

The AGASA experiment has reported an excess near the Galactic center
at energies around 1 EeV \cite{AKaniso}. 
To check this claim, we have prepared a
special low-energy set of events observed by the TASD,
and constructed the event density map averaged
over the circles of 20$^\circ$ centered on the 1$^\circ$$\times$1$^\circ$ grid.
The background was estimated by the time-swapping method.
No significant excesses or deficits were found \cite{TAaniso}.

The result of LSS correlation was done by taking 2MASS Galaxy Redshift
Catalog (XSCz) as LSS mass distribution, proportional to 
ultra high energy cosmic ray source distribution.
The cosmic ray propagation from sources to the Earth is taken account of 
the energy attenuation processes under the assumption that the primary
particles are protons. The arrival directions were smeared
with the 2D Gaussian function of the certain smearing angular width.
The map of the predicted flux was compared to the
sky distribution of the observed events 
(See the right of Figure~\ref{fig:SDaniso}).
At a given smearing angle, the result of the test is
the p-value that shows how likely it is that the cosmic ray
distribution follows the one expected in a given model (LSS or isotropy). 
At low energies E $>$ 10 EeV, the observed data are compatible with
isotropy and not compatible with the structure model unless
the smearing angle is larger than 20$^\circ$.
At high energies E $>$ 57 EeV, the behavior is different. 
The observed data are compatible with the structure
model but incompatible with the isotropic distribution at
the 3$\sigma$ C.L. (pre-trial), for most values of the smearing
angle \cite{TAaniso}.

\section{TA Energy Extended Experiment}

\begin{figure}[ht]
\centering
\includegraphics[width=7cm]{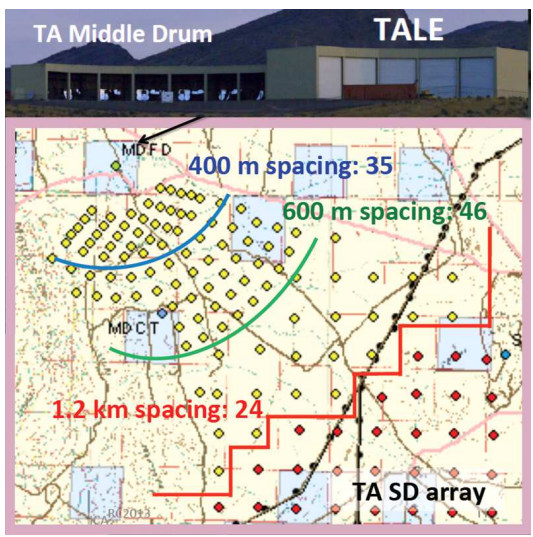}
\caption{TALE site over view.}
\label{fig:TALE}
\end{figure}

The TA Low Energy extension (TALE) project consists
of a set of detectors to be added to TA which will lower the
energy threshold of the experiment to about 10$^{17}$eV. 
The TALE covers area between north west edge of SD array 
and MDFD. (See Figure~\ref{fig:TALE})
The TALE FD consists of 10 new telescopes
which cover elevation angles between 30$^\circ$ and 57$^\circ$.
The TALE SD will consists of a hundred SDs 
which include two stage of spacing infill array and additional 
array to original TASD spacing \cite{TALE}.

The TALE detectors will operate as one experiment with those of TA. 
Cross calibrations of the energy scale and resolution, pointing accuracy, 
and $X_{max}$ reconstruction will be carried out 
for each of the TA/TALE detectors. 
This TA/TALE experiment will have the widest energy range 
in the ultra high energy field. 
They will operate with a single energy scale and will observe cosmic rays
seamlessly from below 10$^{17}$eV to above 10$^{20}$eV.
The physics aims of TALE are to study the second knee
of the cosmic ray spectrum, the galactic-extragalactic transition,
the iron knee and the evolution parameter of the cosmic ray sources. 
In addition, we will characterize cosmic ray showers at 10$^{17}$eV 
to compare with LHC measurements at the equivalent energy of center of momentum.
TALE detectors were partially installed and are operating. 
We are proposing full construction.

The Non-Imaging CHErenkov array (NICHE) is the plan of low energy extension 
for TALE, whose energy range down to 10$^{16}$eV for 200m spacing. 
The main purpose is composition study by Cherenkov pulse-width, 
working with TALE \cite{NICHE}.
 
TA High Energy extension (TAx4) is the plan mainly for anisotropy study.
If the ultra high energy cosmic rays which have energy over GZK are proton, 
their sources are expected within 50 Mpc. 
TAx4 consists of one FD station which have 14 telescopes from HiRes-II and
500 SDs which deployed 3 times larger area with larger spacing than TASD.
Totally, TAx4 SD effective area for GZK energy region can be 4 times larger
than current TASD \cite{TA4}.
We are proposing this construction.

\section{TA Affiliate Experiment}

There are some TA affiliate experiments.

The aim of the EUSO-TA project is to install a prototype of the 
JEM-EUSO telescope at BRFD, and perform observations of ultraviolet 
light generated by cosmic ray air-showers and artificial sources.
This telescopes field of view covers CLF and ELS beams \cite{EUSOTA}.

The Telescope Array Radar (TARA) project will utilize a bistatic radar 
technique to detect radar echoes from the ionization trails 
of ultra high energy cosmic rays as they pass through the Earth's atmosphere.
This method of observing cosmic rays is unproven. 
The effective area of TARA is crossing the TA site. 
TARA will provide confirmation of the radar echo detection of 
ultra high energy cosmic rays \cite{TARA}.

TA Lightning Mapping Array (TALMA) is the project to install 
LMA \cite{LMA} around TASD.
LMA can measure the origin of sferics in the lightning process 
with high resolution of time and position. 
The reason of starting this project is that the TASD detected the bursts of 
the high energy radiation shower events. 

\subsubsection*{Burst events observed by TASD, correlated with lightning}

The TASD observed some short time bursts of air-shower like events.

The definition of this burst is more than 3 events within 1 ms.
There are 10 bursts in five-years SD data.
The expectation of burst in five years is less than $10^{-4}$
by chance coincidence of single events.
There is no selection by position but all these events
for each bursts are very localized, within around 1 km radius.
And some events of 5 bursts in 10 bursts were reconstructed
by air-shower reconstruction.

We checked these bursts with Vaisala lightning database.
This database comes from U.S. National Lightning Detection Network (NLDN),
covering whole TASD site.
NLDN detects lightning by multi-position antennas of low and very low 
frequency band and derives lightning information 
by radio arrival timings and waveforms \cite{NLDN}.

We checked the correlation between 5 reconstructed bursts 
and lightning by timing.
4 bursts in 5 were correlated with lighting within 1 ms.
There is no selection by position but all correlated lightnings
are located in the vicinity of burst air-shower events.
Therefore, TASD burst events is clearly correlated with lightning.

\section{Summary}

Significance of the five-year SD spectrum suppression is 5.7$\sigma$.
Mass composition is consistent with proton.
Arrival direction at low energy seems consistent with isotropic distribution,
at high energy seems inconsistent with isotropic distribution.
Some energy extended experiments and affiliate experiments 
are planned and going on.

\Acknowledgements

The Telescope Array experiment is supported by the Japan
Society for the Promotion of Science through Grants-in-Aids 
for Scientific Research on Specially 
Promoted Research (21000002) ``Extreme Phenomena in the Universe Explored
by Highest Energy Cosmic Rays'' and for Scientific Research
 (19104006), and the Inter-University Research Program of
the Institute for Cosmic Ray Research; by the U.S. National
Science Foundation awards PHY-0307098, PHY-0601915, PHY-0649681,
PHY-0703893, PHY-0758342, PHY-0848320, PHY-1069280, and PHY-1069286; 
by the National Research Foundation
of Korea (2007-0093860, R32-10130, 2012R1A1A2008381, 2013004883); by
the Russian Academy of Sciences, RFBR grants 11-02-01528a 
and 13-02-01311a (INR), IISN project No. 4.4509.10 and Belgian
Science Policy under IUAP VII/37 (ULB). The foundations
of Dr. Ezekiel R. and Edna Wattis Dumke, Willard
L. Eccles and the George S. and Dolores Dore Eccles all
helped with generous donations. The State of Utah supported
the project through its Economic Development Board, and the
University of Utah through the Office of the Vice President
for Research. The experimental site became available through
the cooperation of the Utah School and Institutional Trust
Lands Administration (SITLA), U.S. Bureau of Land Management,
and the U.S. Air Force. We also wish to thank the people
and the officials of Millard County, Utah for their steadfast
and warm support. We gratefully acknowledge the contributions
from the technical staffs of our home institutions. An
allocation of computer time from the Center for High Performance
Computing at the University of Utah is gratefully acknowledged.\\

The lightning data on this report was served from Vaisala Inc.
We appreciate Vaisala's academic research policy.

\end{document}

%% file: econfmacros.tex
\def\beq{\begin{equation}}
\def\eeq#1{\label{#1}\end{equation}}
\def\eeqn{\end{equation}}

\def\beqa{\begin{eqnarray}}
\def\eeqa#1{\label{#1}\end{eqnarray}}
\def\eeqan{\end{eqnarray}}

\let\bar=\overbar

\def\Dslash{\not{\hbox{\kern-4pt $D$}}}
\def\dslash{\not{\hbox{\kern-2pt $\del$}}}

\def\msb{{\bar{\ssstyle M \kern -1pt S}}}

